\documentclass[aps,prd,twocolumn,amsmath,amssymb,showpacs]{revtex4}
\usepackage{graphicx,slashed,cancel,bbm,hyperref,framed,mathrsfs}
\usepackage[T1]{fontenc}
\newcommand{\be}{\begin{equation}}
\newcommand{\ee}{\end{equation}}
\newcommand{\bea}{\begin{eqnarray}}
\newcommand{\eea}{\end{eqnarray}}

\newcommand{\D}{\mathrm{d}}
\newcommand{\E}{\mathrm{e}}
\newcommand{\I}{\mathrm{i}}
\newcommand{\Det}{\mathrm{Det}}
\newcommand{\Tr}{\mathrm{Tr}}
\newcommand{\Lag}{\mathcal{L}}
\newcommand{\V}{\mathcal{V}}
\newcommand{\Z}{\mathcal{Z}}

\newcommand{\four}{\circ}
\newcommand{\five}{\bullet}

\newcommand{\old}{{\varepsilon_\mathrm{old}}}
\newcommand{\new}{{\varepsilon_\mathrm{new}}}

\newcommand{\g}{\mathfrak{g}}
\newcommand{\fg}{\bar{\mathfrak{g}}}

\begin{document}
\title{Thermal worldline holography}
\author{Dennis D.~Dietrich}
\affiliation{Institut f\"ur Theoretische Physik, Goethe-Universit\"at, Frankfurt am Main, Germany}
\affiliation{Arnold Sommerfeld Center, Ludwig-Maximilians-Universit\"at, M\"unchen, Germany}
\begin{abstract}
For a quantum field theory over four-dimensional Minkowski space at zero temperature worldline holography states, that it can be expressed as a field theory of its sources over five-dimensional AdS space to all orders in its elementary fields, the fifth dimension being Schwinger's proper time of the worldline formalism. For the finite temperatures studied here worldline holography yields either a thermal AdS space or an AdS black hole as five-dimensional manifolds. Comparing the values of the five-dimensional action for the two alternatives  does not predict a phase transition as a function of the temperature. This absence is crucially linked to the used coordinates, and worldline holography predicts the Fefferman-Graham form. For these coordinates the only way to tilt the scales in favour of one or the other spacetime is to switch between fermionic or bosonic elementary matter. 
\end{abstract}
\pacs{
11.25.Tq 
12.40.Yx 
}
\maketitle

\section{Introduction}

Strong interactions offer a rich phenomenology, but oftentimes are not accessible with known analytic methods. Holographic techniques promise further analytic understanding. They have been used widely in quantum chromodynamics (QCD) \cite{Erlich:2005qh,Karch:2006pv,Polchinski:2000uf}, in physics beyond the standard model \cite{Hong:2006si,Dietrich:2008ni}, condensed matter systems \cite{Sachdev:2011wg}, and in the context of the Schwinger effect \cite{Sato:2013dwa,Gorsky:2001up,Dietrich:2014ala}. The basis for holography is the conjectured AdS/CFT correspondence \cite{'tHooft:1973jz,Maldacena:1997re} and extensions thereof. None of the presently known instances, however, features a particle content that is realised in nature. For this reason, extrapolations of exactly known correspondences are studied; bottom-up AdS/QCD approaches, for example, reproduce the hadron spectrum of QCD remarkably well \cite{Karch:2006pv,Da Rold:2005zs}. Regrettably, they are not derived from first principles. Therefore, it is crucial to comprehend for which reasons and under which circumstances they amount to a reasonable approximation and which of their characteristics are robust. In particular the role of the extra dimension must be elucidated \cite{deTeramond:2008ht}.

Against this backdrop, worldline holography \cite{Dietrich:2014ala,Dietrich:2016fby,Dietrich:2015oba,Dietrich:2013kza,Dietrich:2013vla} demonstrates that a quantum field theory over four-dimensional Minkowski space can be expressed as a field theory for its sources over five-dimensional anti-de Sitter (AdS) space to all orders in the elementary fields; matter and gauge. Schwinger's proper time of the world line formalism \cite{Strassler:1992zr} naturally takes the role the extra-dimension \footnote{In the worldline formalism one can also derive the Bern-Kosower formalism \cite{Bern:1991aq} without recourse to string theory \cite{Strassler:1992zr,Strassler:1993km}. The use of the worldline formalism also offers a link \cite{Dietrich:2007vw} to the Gutzwiller trace formula \cite{Gutzwiller:1971fy}, which describes quantum systems through classical attributes, as do holographic computations. \cite{Armoni:2008jy} use holography in parallel with the worldline formalism, but do not link Schwinger's proper time to the radial AdS dimension or to anything that follows from there.}.
(The analogous statement also holds for the non-relativistic case \cite{Son:2008ye}.)

Holography at finite temperature is an intensely studied field, for example, in the context of heavy-ion collisions \cite{CasalderreySolana:2011us}. The present paper is concerned with the generalisation of worldline holography to finite temperature:  In thermal quantum field theory the (Euclideanised) time direction is compactified with the inverse temperature $\beta$ as period \cite{Bloch:1932}. This change in topology of the four-dimensional space impacts the geometry of the resulting five-dimensional space. 
In worldline holography (and similarly in holographic approaches in general) the four-dimensional spacetime is the surface of a five-dimensional space at small values of Schwinger's proper time (generally, at one end of the holographic fifth dimension).
In this sense at zero temperature, four-dimensional Minkowski space belongs to five-dimensional AdS space. After compactifying the temporal direction, the fifth-dimensional extrapolation can obviously be the corresponding five-dimensional thermal AdS space;
it, however, can also be a five-dimensional AdS black hole \cite{Hawking:1982dh,Witten:1998zw}. A transition between the two solutions as the temperature is varied is interpreted as confinement-deconfinement transition \cite{Witten:1998zw,Rey:1998bq}; the thermal AdS space corresponding to the confined phase and the black hole solution to the deconfined one. ~ Aforesaid periodicity can be transferred directly to the worldline formalism \cite{McKeon:1992if}, and below we shall see that at finite temperature the admissible five-dimensional spaces in worldline holography are again the thermal AdS space and the AdS black hole.

In Section \ref{sec:zero} we display aspects of worldline holography at zero temperature required as reference for the changes brought about by admitting a finite temperature, and which are discussed in \ref{sec:finite}. The final section concludes the paper.

\section{Worldline holography}

\subsection{Zero temperature \label{sec:zero}}

\subsubsection{Volume elements}

Here, we recapitulate the principles of worldline holography \cite{Dietrich:2014ala,Dietrich:2016fby,Dietrich:2015oba,Dietrich:2013kza,Dietrich:2013vla} exclusively for a vector source $V$ so as not to obscure the underlying structure. (It is straightforward to include general sources.) For the same reason, we show all expressions for scalar elementary matter, which saves us from displaying another functional Grassmann integral or global trace. (We are going to comment on the fermionic result below.) Moreover, we also keep implicit global finite dimensional traces and the path ordering required by the generally non-Abelian nature of the gauge (colour) and flavour groups. The correlators of the sources $V$ are encoded in the generating functional
\begin{align}
Z_{\varepsilon}=\langle\E^w\rangle=\int[\D G]\E^{w-\frac{i}{2e^2}\int\D^4x\,\mathrm{tr}\, G_{\mu\nu}^2}.
\label{eq:genfun}
\end{align}
The functional integral runs over the gauge field $G_\mu$.  $G_{\mu\nu}=\partial_\mu G_\nu-\partial_\nu G_\mu-\I[G_\mu,G_\nu]_-$. Here,
\begin{align}
w
=
-\frac{1}{2}\ln\Det(\mathbbm{D}^2)
=
-\frac{1}{2}\Tr\ln(\mathbbm{D}^2),
\label{eq:scalar}
\end{align}
where 
$\mathbbm{D}=\partial-\I\mathbbm{V}$,
\begin{align}
\mathbbm{V}=G+V= G^jT^j_\mathrm{colour}\mathbbm{1}_\mathrm{flavour}+V^jT^j_\mathrm{flavour}\mathbbm{1}_\mathrm{colour},
\end{align}
 and the $T^j_\mathrm{colour}$ ($T^j_\mathrm{flavour}$) are the generators of $\mathcal{G}_\mathrm{colour}$ ($\mathcal{G}_\mathrm{flavour}$). 
The worldline formalism \cite{Strassler:1992zr} translates $w$ into \cite{Dietrich:2014ala,Dietrich:2016fby,Dietrich:2015oba,Dietrich:2013kza,Dietrich:2013vla}
\begin{align}
w
=
&\int \D^4x_0\int_{\varepsilon>0}^\infty\frac{\D T}{2T^3}\,\Lag
\label{eq:w}
\equiv\\\equiv
&\int\D^5x\,\sqrt{|g|}\,\Lag ,
\label{eq:wg}
\end{align}
with the Lagrangian density 
\be
\Lag
=
\frac{2\mathcal{N}}{(4\pi)^2}
\int_\mathrm{P}[\D y]\;\E^{-\int_0^T\D\tau[\frac{\dot y^2}{4}+\I \dot y \cdot\mathbbm V(x_0+y)]} .
\label{eq:lag}
\ee
$\sqrt{|g|}$
is the volume element for an AdS$_5$ metric with the parametrisation
\be
\D s^2\overset{g}{=}-\frac{\D T^2}{4T^2}+\frac{\D x_0\cdot\D x_0}{T}.
\label{eq:patch}
\ee
`$\cdot$' indicates a contraction with the flat four-dimensional metric $\eta_{\mu\nu}$. $T$ stands for Schwinger's proper time, which was first introduced to exponentiate the logarithm in \eqref{eq:scalar}, which gives rise to one inverse power of $T$. The proper-time regularisation $T\ge\varepsilon>0$ coincides with the UV-brane regularisation in holography. The functional trace Tr (after Wick rotation) is expressed as a path integral \eqref{eq:lag} over all closed paths, $y^\mu(0)=y^\mu(T)$. The $\D^4x_0$ integral translates the paths to every event in spacetime, $x^\mu\equiv y^\mu+x_0^\mu$. [$x_0\neq x_0(\tau)$ is split off from the path integral, because it is the zero mode of the operator $\partial^2_\tau$ ($\dot y\equiv\partial_\tau y$) in the kinetic term of the worldline action. $x_0$ can, for example, be chosen as the starting point of the paths, then $y^\mu(0)=0=y^\mu(T)$; or the `centre-of-mass', then $\int_0^T\D\tau\, y=0$.] $x_0$ and $T$ together span the AdS$_5$ space. In the course of the derivation of the path integral in \eqref{eq:lag} there arises a mismatch in the number of intergrations over four-dimensional momentum and position space, respectively. This leads to two additional inverse powers of $T$. The total of three inverse powers make up the AdS$_5$ volume element. 
That the AdS$_5$ metric materialises at this points is an echo of the conformal symmetry SO(4,2) of 3+1 dimensional Minkowski space, which is also the symmetry group of AdS$_5$. 
The $\mathbbm V$ dependent piece in \eqref{eq:lag} amounts to a Wilson loop $\E^{-\I\oint\D y\cdot\mathbbm V}$, which is manifestly invariant under the local transformations $\mathbbm{V}^\mu\rightarrow\Omega[\mathbbm{V}^\mu+\I\Omega^\dagger(\partial^\mu\Omega)]\Omega^\dagger$, where $\Omega\in\mathcal{G}_\mathrm{colour}\times\mathcal{G}_\mathrm{flavour}$. Thus, hidden local symmetry \cite{Bando:1984ej} emerges.\footnote{For the subset of diagrams where all sources are connected to the same matter loop, most of the following steps are not needed, and there are arguments \cite{Okubo:1963fa,Shifman:1978bx,Dietrich:2012un} that such diagrams could be dominant.}

$e^w$ in \eqref{eq:genfun} features powers of $w$, and $w^n$ contains $n$ integrations over $\D^5 x_j$,
\be
w^n
=
\prod_{j=1}^n
\int\D^4x_j\int_{\varepsilon>0}^\infty\frac{\D T_j}{2T_j^3}\,\Lag_j.
\label{eq:wn}
\ee
Let us first substitute absolute, $x_0$, and dimensionless relative coordinates $\hat\Delta$, such that the Jacobian equals $T^{2(n-1)}/2$ for the sake of concreteness,

\begin{align}
\int\D^{4n}x=\int\D^4x_0\D^{4(n-1)}\Delta=\frac{T^{2(n-1)}}{2}\int\D^4x_0\D^{4(n-1)}\hat\Delta.
\end{align}

Subsequently, we transform to an overall proper time $T$ and fractions $t_j$ thereof, using
\be
1=\int\D T\,\delta\Big(T-\sum_{j=1}^n T_j\Big)\prod_{j=1}^n\Big[\int\D t_j\,\delta\Big(t_j-\frac{T_j}{T}\Big)\Big].\label{eq:choco}
\ee
The $\delta$-s in the product can be used to carry out the $\D T_j$ integrations, which leaves behind a factor of $T^n$. With the $\D T_j$ integrations gone, the first $\delta$ can be rewritten according to 
$\delta(T-\sum_{j=1}^n T_j)=\delta(1-\sum_{j=1}^n t_j)/T$. Gathering all the powers of $T$: from $\D^{5n}x$, $T^{-3n}$; from the aforementioned Jacobian, $T^{2(n-1)}$; and from evaluating \eqref{eq:choco} in \eqref{eq:wn}, $T^{n-1}$; gives a total of $T^{-3}$. This factor makes up the volume element $\sqrt{|g|}$ in
\begin{align}
w^n
=&
\int d^5x\sqrt{|g|}
\int\D^{4(n-1)}\hat\Delta
\nonumber\times\\&\times
\prod_{j=1}^n\int_{\frac{\epsilon}{T}}^{1-\frac{\epsilon}{T}}\frac{dt_j}{2t_j^3}\,\Lag_j
\times\delta\Big(1-\sum_{k=1}^nt_k\Big).
\end{align}
Consequently, also $w^n$ possesses the form of an action over AdS$_5$.

\subsubsection{Contractions with the metric}

In the next step we have to verify that all indices on fields and gradients are contracted with AdS$_5$ metrics. According to \eqref{eq:patch}, the difference between the AdS and the flat metrics are special powers of $T$. In order to isolate them, we convert to integration variables without mass dimension, $\hat y_j=y_j/\sqrt{T_j}$ and $\hat y_j=y_j/\sqrt{T_j}$. Then $\mathcal{L}_j$ takes the form
\begin{align}
\label{eq:lagj}
&\Lag_j
=
\frac{2\hat{\mathcal{N}}}{(4\pi)^2}
\int_\mathrm{P}[\D\hat y_j]
\times\\\nonumber
&\E^{-\int_0^1\D\hat\tau_j[\frac{(\partial_{\hat\tau_j}\hat y_j)^2}{4}+\I\sqrt{t_jT}(\partial_{\hat\tau_j}\hat y_j)\cdot\E^{\eta\sqrt{T}(\sqrt{t_j}\hat y_j+\widehat{x_j-x_0})\cdot \partial_{x_0}}\mathbbm V(x_0)]}.
\end{align}
($\widehat{x_j-x_0}$ depends only on $\hat\Delta$ not $x_0$.)
This shows that every gradient $\partial_{x_0}$ and every field $\mathbbm V$ is accompanied by exactly one power of $\sqrt{T}$.

This fact is not changed by carrying out the integral over the gauge field $G$: The gauge-field average of $\prod_{j=1}^n\Lag_j$ reads
\begin{align}
\nonumber
&\Big\langle\prod_{j=1}^n\Lag_j\Big\rangle
=
\bigg(\frac{2\hat{\mathcal{N}}}{(4\pi)^2}\bigg)^n
\prod_{j=1}^n\int_\mathrm{P}[\D\hat y_j]
\nonumber\\&\nonumber
\E^{-\int_0^1\D\hat\tau_j [\frac{(\partial_{\hat\tau_j}\hat y_j)^2}{4}+\I\sqrt{t_jT}(\partial_{\hat\tau_j}\hat y_j)\cdot\E^{\eta\sqrt{T}(\sqrt{t_j}\hat y_j+\widehat{x_j-x_0})\cdot \partial_{x_0}}V(x_0)]}
\\&
\Big\langle
\prod_{l=1}^n\E^{\I\oint\D y_l\cdot G(x_l+y_l)}
\Big\rangle.
\label{eq:aint}
\end{align}
After the $[\D G]$ integration has been carried out the average on the last line of \eqref{eq:aint} does not depend on the gauge field anymore, is homogeneous of degree 0 in the dimensionfull variables $\{x_j+y_j|j=1\dots n\}$, and does not depend on $x_0$, because of translational invariance. Thanks to the aforementioned homogeneity it does not change its value if we replace $x_j+y_j\rightarrow\hat x_j+\hat y_j~\forall j$ while preserving $x_0$, on which it does not depend anyhow. Moreover, said average also does not depend on $T$, as neither the Wilson lines nor the integration weight were depending on $T$.

Thus, after executing the $[\D\hat y]$ and $\D\hat\Delta$ integrations as well, the expression only depends on $T\eta^{\mu\nu}W_\mu W_\nu$, for all combinations of $W\in\{\partial_{x_0},V\}$. These can be reexpressed as $g^{\mu\nu}W_\mu W_\nu$.
Therefore, all contractions of indices are carried out with (inverse) AdS metrics and no other factors of $T$ occur beyond that.

\subsubsection{5d dependence}

\cite{Dietrich:2014ala,Dietrich:2015oba,Dietrich:2013kza} had identified the fifth-components of gradients and fields as being due to an induced Wilson (gradient) flow \cite{Luscher:2009eq} fixed by a variational principle. \cite{Dietrich:2016fby} showed that the exact same result is found by demanding the independence of the effective action $Z_\varepsilon$ from the ultraviolet proper-time regulator $\varepsilon$, which comes down to a Wilson-Polchinsky renormalisation condition. 

In order to see this consider \eqref{eq:genfun} after integrating out all proper-time fractions $t_j$ and $\hat\tau_j$ as well as dimensionless relative coordinates $\hat\Delta$, expanded in powers of the covariant derivative $D=\partial-\I V$,
\begin{align}
Z_\varepsilon=\iint_\varepsilon^\infty\D^5x\sqrt{|g|}\sum_n \#_n (g^{\four\four})^n(D_\four)^{2n}.
\label{eq:genfuncser}
\end{align}
This representation is manifestly locally invariant under the transformation $V^\mu\rightarrow\Omega[V^\mu+\I\Omega^\dagger(\partial^\mu\Omega)]\Omega^\dagger$, $\Omega\in\mathcal{G}_\mathrm{flavour}$ \cite{Shifman:1980ui,Schmidt:1993rk} as is the interaction part in \eqref{eq:lag} $\E^{-\I\oint\D y\cdot V}$, a Wilson loop. (This brings about hidden local symmetry \cite{Bando:1984ej}.) The proper-time regularisation leaves this symmetry intact. Here, the $\#_{n_\partial,n_V}$ are dimensionless numerical coefficients. `$\four$' indicates that the five-dimensional (inverse) metric $g$ is only used for contractions in four dimensions. The addends in \eqref{eq:genfuncser} symbolise all occurring combinations of contractions, which may be antisymmetrised. Particularly, \eqref{eq:genfuncser} is no differential operator, i.e., no partial derivatives act to the right. The leading term ($n=2$), for example, is $\propto g^{\mu\kappa}g^{\nu\lambda}[D_\mu,D_\nu]_-[D_\kappa,D_\lambda]_-$.

The value of the proper-time regulator $\varepsilon>0$ in \eqref{eq:genfuncser} has (a priori) no physical meaning. Therefore, $Z_\varepsilon$ must not depend on the value of $\varepsilon$, or $Z_\old\overset{!}{=}Z_\new$ for $\old\neq\new$. In order to see what has to be done to achieve this independence let us try to undo the change $\old\rightarrow\new$ in
\begin{align}
Z_\new
=
\int\D^4x_0\int_\new^\infty&\frac{\D T}{2T^3}\sum_n\#_n
(T\eta^{\four\four})^n(D_\four)^{2n}.
\end{align}
This is to be accomplished for all configurations of $V$. Therefore, the independence must be enforced order by order, i.e., $\forall~n$ separately. Thus, we have to change the integration bound without changing the integrand. Rescaling the integration variables globally,
\begin{align}
T\rightarrow c_T T~~~~~\&~~~~~x_0\rightarrow c_x x_0
\end{align}
and consequently
\begin{align}
\partial_{x_0}\rightarrow c_x^{-1}\partial_{x_0}
\end{align}
necessitates a simultaneous rescaling of $V$,
\begin{align}
V\rightarrow c_x^{-1}V,
\end{align}
because it appears only inside the covariant derviative in combination with the partial derivative. Hence,
\begin{align}
&Z_\new
=\\=&
\int\D^4x_0\int_{c_T\new}^\infty\frac{\D T}{2T^3}\sum_n\#_n
c_x^{4-2n}c_T^{n-2}
(T\eta^{\four\four})^n(D_\four)^{2n}.
\nonumber
\end{align}
We restore the original integration bound with the choice $c_T=\old/\new$. The invariance of the integrand is achieved for $c_x=c_T^{1/2}$. This comes at the cost of endowing the source $V$ with a dependence on the regulator, which is a fifth-dimensional quantity,
\begin{align}
\old V(x_0;\old)=\new V(x_0;\new).
\end{align}

~\\

Then again, $Z_\varepsilon$ already has the form of an action over AdS$_5$ \cite{Dietrich:2014ala,Dietrich:2013kza,Dietrich:2013vla,Dietrich:2015oba,Dietrich:2013kza}, and the isometries of AdS$_5$ coincide with the conformal group of Mink$_4$, including scale-invariance. Therefore, the independence from the value of $\varepsilon$ is also achieved by incorporating the fifth components of gradients and sources into \eqref{eq:genfuncser}. There were, however, no fifth-dimensional polarisations in the four-dimensional theory. We are allowed to remove them by the gauge condition $\V_T=0$, if a corresponding gauge symmetry exists. Since \eqref{eq:genfuncser} is invariant under four-dimensional local flavour transformations,
\begin{align}
\label{eq:calz}
\Z
=
\iint_\varepsilon^\infty\D^5x\sqrt{|g|}\sum_n \#_n (g^{\five\five})^n(\mathbbm{D}_\five[g])^{2n}.
\end{align}
is invariant under five-dimensional ones. Here `$\five$' indicates the full five-dimensional contraction, and $\mathbbm{D}[g]$ stands for the flavour and AdS covariant derivative. \eqref{eq:calz} is also conformally invariant if $\V(x_0,T)$ transforms like a five-dimensional vector. ($\V$ does not depend explicitly on $\varepsilon$.)

Even when imposing the $\V_T=0$ gauge already at this point, the needed scale invariance is still manifest, because scale transformations do not mix tensor components, while the special conformal transformations do. 
The full conformal invariance, however, would still guaranteed modulo a subsequent local flavour transformation.

$\Z$ as such is only a functional of arbitrary source configurations. Its genuine significance as an action defining its field theory is by virtue of the configuration(s) $\breve\V$ it marks as saddle points. 
The boundary condition
\begin{align}
\breve\V_\mu(x_0,T=\varepsilon)=V_\mu(x_0)
\label{eq:inicond}
\end{align}
transfers the four-dimensional polarisations to the five-dimensional field $\V$ and bestows it with the same normalisation like $V$, which is the source for {\sl once} the vector current.
[Consistently, the worldline expressions follow a Wilson (gradient) flow equation with this boundary condition\cite{Dietrich:2014ala,Dietrich:2015oba,Dietrich:2013kza}.]

In sum, the cutoff independent effective action is $\Z$ evaluated on the saddle point with \eqref{eq:inicond} and $\breve\V_T=0$,
\begin{align}
\breve\Z
=
\iint_\varepsilon^\infty\D^5x\sqrt{|g|}\sum_n \#_n (g^{\five\five})^n(\breve{\mathbbm{D}}_\five[g])^{2n}.
\end{align} 

With small $T$ corresponding to small distances, \eqref{eq:inicond} positions the bare source configuration at the ultraviolet end of the fifth dimension. In conjunction with the requirement that the effective action do not depend on the value of the unphysical UV regulator $\varepsilon$,  
\be
\varepsilon\partial_\varepsilon \ln Z_\varepsilon\overset{!}{=}0,
\label{eq:renorm}
\ee
makes this a Wilson-Polchinsky renormalisation condition \cite{Wilson:1971bg}.

Lastly, this is also the computation carried out in the holographic models \cite{Karch:2006pv,Erlich:2005qh,Polchinski:2000uf,Maldacena:1997re}. There the four-dimensional effective action is equated with the five-dimensional action evaluated on its saddle point. Worldline holography identifies Schwinger's proper time as the fifth dimension \cite{Dietrich:2014ala,Dietrich:2013vla,Dietrich:2015oba,Dietrich:2013kza} and the fifth-dimensional profile of the sources as solution to a renormalisation group equation \eqref{eq:renorm}.

\subsubsection{Spin-2}

In this paper, we will be mostly concerned with spin-2 backgrounds. For them one can devise a generally covariant expansion analogous to the vector case. For a given four-dimensional metric $\g$ one has
\begin{align}\nonumber
Z_\varepsilon
&=
\int_\varepsilon^\infty\frac{\D T}{2T^3}\int \D^4x_0\sqrt{|\g|}\sum_n \#_n (T\g^{\four\four})^n(\nabla_\four[\g])^{2n}
=\\&=
\iint_\varepsilon^\infty\D^5x\sqrt{|\fg|}\sum_n \sharp_n (\fg^{\four\four})^n(\nabla_\four[\g])^{2n},
\label{eq:dewitt}
\end{align}
where $\nabla[\g]$ represents the Levi-Civita connection for $\g$. [The same remarks for reading this expression apply as before, i.e.,  once more, \eqref{eq:dewitt} is not a differential operator; there are no partial derivatives acting to the right; the expansion is to symbolise all occurring combinations; the commutator of two covariant derivatives, for example, gives the Riemann tensor.]
On the second line, $\fg$ stands for the five-dimensional Fefferman-Graham \cite{Fefferman:1985} embedding of $\g$,
\begin{align}
\D s^2\overset{\fg}{=}\natural\Big(\frac{\D T^2}{4T^2}+\frac{\g_{\mu\nu}\D x^\mu\D x^\nu}{T}\Big),
\label{eq:fg}
\end{align}
and $\sharp_n=\#_n\natural^{n-5/2}$.
Hence, also for spin-2, the four-dimensional effective action takes the form of a five-dimensional field theory modulo fifth components. Since $\g$ is not required to be flat, $\fg$ need not be an AdS space. The deviations $h$ in an expansion of $\g$ around flat space, $\g_{\mu\nu}=\eta_{\mu\nu}+h_{\mu\nu}$, do form a field theory over AdS$_5$. The $h_{\mu\nu}$ are the sources to the four-dimensional stress-energy tensor.

In order to ensure the independence from $\varepsilon$ we can evaluate the five-dimensional completion of \eqref{eq:dewitt}, 
\begin{align}
\Z
=
\iint_\varepsilon^\infty\D^5x\sqrt{|\fg|}\sum_n \sharp_n (\fg^{\five\five})^n(\nabla_\five[\fg])^{2n}
\end{align}
on the saddle point with the boundary condition 
\begin{align}
\breve\fg_{\mu\nu}(x_0,T=\varepsilon)=\frac{\natural}{\varepsilon}\g_{\mu\nu}(x_0)
\label{eq:bcg}
\end{align}
and the gauge condition
\begin{align}
\breve h_{TN}\overset{!}{=}0\,\forall N
\label{eq:gcg}
\end{align}
which eliminates deviations with fifth-dimensional polarisations \cite{Dietrich:2016fby}.

The $\#_n$ are the known DeWitt-Gilkey-Seeley coefficients \cite{DeWitt:1965}. For the problem at hand, the first two correspond to a negative cosmological constant and an Einstein-Hilbert term,
\begin{align}
\Z_\varphi=\frac{1}{6(4\pi)^2}\iint_\varepsilon^\infty\D^5x\sqrt{|\fg|}(R[\fg]+6),
\label{eq:eh}
\end{align} 
at one-loop.
Therefore, the saddle-point equations are the Einstein equations and possess an AdS$_5$ solution with the squared AdS curvature radius 
\begin{align}
\natural=\frac{(5-1)(5-2)}{6}=2.
\end{align}
such that for \eqref{eq:bcg} and \eqref{eq:gcg}
\begin{align}
\D s^2\overset{\breve\fg}{=}2\Big(\frac{\D T^2}{4T^2}+\frac{\D x_0\cdot\D x_0}{T}\Big)
\label{eq:fgsad}
\end{align}
and 
\begin{align}
R[\breve{\fg}]=-\frac{5(5-1)}{\natural}=-\frac{5(5-1)}{(5-1)(5-2)}6=-10
\end{align}
such that
\begin{align}
\breve\Z_\varphi=-\frac{16\sqrt{2}}{6(4\pi)^2}\iint_\varepsilon^\infty\D^5x\sqrt{|g|}.
\label{eq:eh}
\end{align}
In what follows, we shall study the formalism up to this order. (A space of constant curvature remains a saddle-point solution also at higher $n$, but, generally, with a different curvature radius.)
For any value of the curvature radius the isometries of an AdS$_5$ space are the conformal group over Mink$_4$ (just like Mink$_4$ features Poincar\'e invariance whatever the speed of light) making the value of the AdS radius of secondary importance. As concrete example,
\begin{align}
\Z
=
\iint_\varepsilon^\infty\D^5x\sqrt{|\breve{\fg}|}\sum_n \#_n (\breve{\fg}^{\five\five})^n(\mathbbm{D}_\five[\breve{\fg}])^{2n}.
\end{align}
where $\sharp_{n_\partial,n_V}=\#_{n_\partial,n_V}\natural^{(n_\partial+n_V-5)/2}$, is an identical reexpression for \eqref{eq:calz}, which is independent of $\natural$, because $g$ is. Consistently, the covariant derivatives do not depend on the curvature radius and, accordingly, neither does the (1,3) Riemann tensor.

For fermionic elementary matter instead, the coefficients are \cite{Shapiro:2008sf}
\begin{align}
\Z_\psi=-\frac{1}{3(4\pi)^2}\iint_\varepsilon^\infty\D^5x\sqrt{|\fg|}(R[\fg]+12),
\label{eq:eh}
\end{align} 
at one loop. Consequently,
$
\natural=1,~
$
$
\breve\fg=g,~
$
$
R[\breve{\fg}]=-20,~
$
and
\begin{align}
\breve\Z_\psi=+\frac{20}{3(4\pi)^2}\iint_\varepsilon^\infty\D^5x\sqrt{|g|}.
\label{eq:eh}
\end{align}

\subsection{Finite temperature\label{sec:finite}}

In thermal field theory the (Euclideanised) time direction $t$ is compactified with period of inverse temperature $\beta$ \cite{Bloch:1932}. [As here $T$ already stands for Schwinger's proper time, we will only use $\beta$ for denoting the (inverse) temperature.] The periodicity of the temporal direction can be implemented straightforwardly in the worldline formalism \cite{McKeon:1992if}. Thus, the expressions for $Z$ in \eqref{eq:genfun} and $w$ in \eqref{eq:w} remain unchanged up to the topological constraint
\be
t\equiv t+n\beta~\forall n\in\mathbbm Z.
\label{eq:topconstr}
\ee
As a consequence, the AdS$_5$ metric \eqref{eq:patch} still appears in \eqref{eq:wg}, only that the time direction of the five-dimensional space is compactified accordingly. This is known as thermal AdS space. Powers of $w$ can also be shown to correspond to actions over AdS$_5$ by the same changes of variables as before. One merely has to be careful in determining the integration bounds for the temporal components of the relative coordinates, because of \eqref{eq:topconstr}. Thus, at finite temperature all volume elements in $Z$ are those of (thermal) AdS$_5$ \eqref{eq:patch}.

Accordingly, \eqref{eq:lagj} can also be obtained in the thermal case; the compactification of the spatial dimension does not change the fact that every gradient and every field is accompanied by its own factor of $\sqrt{T}$. 
The gauge-field averaged Wilson loop, obtained by carrying out the $[\D G]$ integration, is homogeneous of degree 0 under the simultaneous rescaling of $\{x_j+y_j|j=1\dots n\}$ and $\beta$. As a consequence, the replacement $x_j+y_j\rightarrow\hat x_j+\hat y_j~\forall j$ 
together with $\beta\rightarrow\hat\beta=\beta/\sqrt{T}$ leaves the value of aforesaid average unchanged.
Hence, after integrating out $[\D\hat y]$ also at finite temperature all factors of $T$ are accounted for by contractions with (inverse) AdS$_5$ metrics, while the gauge field average is, in general, depending on $\hat\beta$ (the period of the $\hat t$ direction). 

Finally, also at finite temperature there persists the induced fifth-dimensional profile of the sources.
The corresponding equations of motion of $V$ are modified by the presence of the scale $\beta$. 
In the remainder of this paper, however, we will concentrate on peculiarities of the thermal five-dimensional geometry.

\subsubsection{The emergent five-dimensional spaces}

There are two different fifth-dimensional extrapolations of four dimensional flat space with one compactified dimension \cite{Hawking:1982dh,Witten:1998zw}, the thermal AdS space already encountered above and the AdS black hole
\be
\D s^2=-\frac{\D T^2}{4T^2}+\frac{\D t^2(1-\frac{T^2}{T^2_h})^2/(1+\frac{T^2}{T^2_h})+ \D\vec x^2(1+\frac{T^2}{T^2_h})}{T}.
\label{eq:patchblack}
\ee
For the AdS black hole \eqref{eq:patchblack} the periodicity in the temporal direction and hence the temperature arises from a regularity condition for the metric at the black-hole horizon \cite{Hartle:1976tp}, $T_h=2(\frac{\beta}{\pi})^2$.
The difference between the thermal AdS and the AdS black-hole metrics vanishes at proper time $T=0$, as does its first derivative with respect to $T$. This is the underlying reason why an extrapolation from four to five dimensions requires additional pieces of information to decide between the two metrics. In fact, \cite{Bautier:2000mz} treats the more general case of extrapolating a general conformally flat metric in $d$ dimensions to a metric in the Fefferman-Graham form \cite{Fefferman:1985} in $d+1$ dimensions \eqref{eq:fg}.
For the present case, \cite{Bautier:2000mz} concludes, that the difference between thermal AdS and AdS black holes is contained in parts of the extrapolated $\g$, which are not fixed by choosing the boundary condition \eqref{eq:bcg}.
The difference is found to always start at $O(T^{d/2}=T^2)$, which is also the case for \eqref{eq:patchblack} relative to \eqref{eq:patch}. [Furthermore, the $T^2$ term is also always traceless when traced with $\g(x,{T\rightarrow0})$ \cite{Bautier:2000mz}, which checks out for \eqref{eq:patchblack} as well.]

Having two saddle-point solutions it remains to find the one which gives the minimal action by putting the solution back into the action. Also the black-hole metric has constant scalar Ricci curvature $-20$, hence the comparison of the values of the actions amounts to the comparison of the spacetime volumes,
\begin{align}
\mathrm{Vol}_5=\iint_\varepsilon^\infty\D^5x\sqrt{|g|}
\end{align}
versus
\begin{align}
\mathrm{Vol}_5=\iint_\varepsilon^{T_h}\D^5x\sqrt{|g|}\Big(1-\frac{T^4}{T_h^4}\Big).
\end{align} 
We find
\be
\frac{\mathrm{Vol}_5-\mathrm{Vol}_5^\mathrm{BH}}{\mathrm{Vol}_3\,\beta}\propto\frac{1}{2T_h^2},
\label{eq:vols}
\ee
where we have taken the limit $\varepsilon\rightarrow 0$. (See Appendix \ref{sec:matching} for a comment on some subtleties concerning the matching of the two spaces.) Therefore, for this setup with  fermionic  elementary matter the AdS black hole would be preferred at every finite temperature. 
The most notable difference for  bosonic  elementary matter is an overall sign on the effective actions. 
As a consequence, there the thermal AdS space would be preferred for all values of the temperature.

The above does not change if we bring back the warp factor $\E^{-m^2T}$, generated by the mass of the elementary matter.
 (For a discussion of the origin of this term in worldline holography see Appendix \ref{sec:mass}.) 
There
\begin{align}\label{eq:volsm}
\frac{\mathrm{Vol}_5-\mathrm{Vol}_5^\mathrm{BH}}{\mathrm{Vol}_3\,\beta}\propto{}&
\frac{1}{4T_h^2}\Big[\Big(1-\frac{2}{m^2T_h}-\frac{2}{m^4T_h^2}\Big)
+\\&+\nonumber
\frac{2}{m^4T_h^2}-m^4T_h^2\mathrm{Ei}(-m^2T_h)\Big]>0,
\end{align}
where Ei stands for the exponential integral. (For more details about the effect of warping see Appendix \ref{sec:warp}.)

Thus, for  fermionic elementary matter  
the AdS black hole, which is usually associated with the unconfined phase in holography \cite{Witten:1998zw,Rey:1998bq}, is preferred. This interpretation is consistent, as for the comparison of the two spacetimes all particle interactions had been switched off. Consistently, the warping due to the mass of the elementary particles also did not change this picture.
To the contrary, for  bosonic elementary matter  the thermal AdS solution is preferred. This solution is commonly associated with the confined phase, but in the absence of interactions, this interpretation does not appear sensible. 

\section{Short summary}

Worldline holography \cite{Dietrich:2014ala,Dietrich:2016fby,Dietrich:2015oba,Dietrich:2013kza,Dietrich:2013vla} makes a link between the Schwinger proper time of the worldline formalism and the fifth dimension in AdS holography, by showing that a quantum field theory over four-dimensional Minkowski naturally becomes a field theory of its sources over five-dimensional AdS space to all orders in the elementary fields; matter and gauge. 
The present paper investigated the effect of finite temperature. Starting with the worldline formalism at finite temperature \cite{McKeon:1992if} and repeating the derivation of the worldline holographic background geometry, one arrives at a field theory over thermal AdS space.  In worldline holography, however, there also occur deviations from aforesaid background geometry. These deviations are sources for the stress-energy tensor. Together with the background they form a spacetime geometry, which is determined by equations of motion. For the given boundary conditions (and especially topology) there arise two solutions. In one the deviations vanish, i.e., we keep the thermal AdS space. The other solution is an AdS black hole. 
Comparing the otherwise unsourced actions of these two geometries 
yields a lower value for the thermal AdS space (AdS black hole)  for bosonic (fermionic)  elementary  matter  of freedom  for all values of the temperature. Particularly this means that there is no phase transition as a function of the temperature. This result crucially depends on which coordinates are chosen for the AdS black hole, but the worldline formalism predicts those in the Fefferman-Graham form, for which, as a consequence, we carried out the analysis. (Without a formalism to choose the coordinates, one could carry out the computation in any coordinate patch, but this leads to ambiguities.) Moreover, these results already arise at the one-loop level, i.e., without interactions, i.e., without any basis for confinement. This is in contrast to the fact that there are situations, where the thermal AdS space is preferred, and it is traditionally interpreted as describing the confined phase.

\section*{Acknowledgments}

D.D.D.~would like to thank
Stan Brodsky,
Guy de T\'eramond,
Luigi Del Debbio,
Hans-G\"unter Dosch,
Gerald Dunne,
Gia Dvali,
Joshua Erlich,
Daniel Flassig,
Christiano Germani,
C\'esar G\'omez,
Alexander Gu{\ss}mann,
Stefan Hofmann,
Paul Hoyer,
Adrian K\"onigstein,
Sebastian Konopka,
Michael Kopp,
Matti J\"arvinen,
Florian Niedermann,
Yaron Oz,
Joachim Reinhardt,
Tehseen Rug,
Ivo Sachs,
Debajyoti Sarkar,
Andreas Sch\"afer,
Robert Schneider,
Karolina Socha,
Philippe Spindel,
Stefan Theisen,
Christian Weiss,
Nico Wintergerst,
and
Roman Zwicky
for discussions. 
The work of the author was supported by the Humboldt foundation and the European Research Council.

\appendix

\section{Matching of the spaces\label{sec:matching}}

The $\D T$ integrals diverge at their lower boundary. Therefore, we had regularised them from the very beginning.  
While the periodicities in the temporal direction at $T=0$ coincide for the same inverse temperature in both spaces, they do not at $T=\varepsilon> 0$. In accordance with the different prefactors of the $\D t^2$ terms in the two line elements, we must choose a periodicity $\beta\times(1-\epsilon^2/T^2_h)/(1+\epsilon^2/T^2_h)^{1/2}$ for the thermal AdS space if the periodicity in the black hole space is $\beta$. Furthermore, due to the different prefactors of the $\D\vec x^2$ addends, also the three volumes (which are formally infinite, because of translational invariance, and must thus be regularised as well) differ at $T=\varepsilon> 0$. Choosing Vol$_3$ for the AdS black hole, the thermal AdS case has a factor of Vol$_3\times(1+\epsilon^2/T^2_h)^{3/2}$. Together this amounts to a factor of Vol$_3\,\beta\times(1-\epsilon^4/T^4_h)$ on the integral for the thermal AdS space. It deviates from the unadjusted result at $O(\varepsilon^4)$. Therefore the adjustment does not contribute in the limit $\varepsilon\rightarrow0$, because the strongest divergence is only $\propto\varepsilon^{-2}$ and $\lim_{\varepsilon\rightarrow0}\varepsilon^4/\varepsilon^2=0$. The divergence $\propto\varepsilon^{-2}$ cancels in the difference of the two integrals.

\section{Massive matter \label{sec:mass}}

For massive elementary matter with mass $m$ in the worldline formalism leads to an additional factor of $\E^{-m^2T}$ in the integrand of \eqref{eq:wg}. In the worldline holographic context we include a scalar source $S$ instead, such that 
\eqref{eq:scalar} becomes
\begin{align}
w
=
-\frac{1}{2}\Tr\ln(\mathbbm{D}^2+S^2),
\end{align}
and, accordingly, \eqref{eq:lag}
\begin{align}
\Lag
=
\frac{2\mathcal{N}}{(4\pi)^2}
\int_\mathrm{P}[\D y]\;\E^{-\int_0^T\D\tau[\frac{\dot y^2}{4}+\I \dot y \cdot\mathbbm V(x_0+y)+S^2(x_0+y)]} .
\end{align}
To the lowest order in gradients and all orders in the scalar source, this contains
\begin{align}
Z_\varepsilon\supset\#\iint_\varepsilon\D^5x\sqrt{|g|}\E^{-TS^2(x_0)},
\end{align}
which in the five-dimensional $\varepsilon$ independent form becomes
\begin{align}
\mathcal{Z}\supset\#\iint_\varepsilon\D^5x\sqrt{|g|}\E^{-\mathcal{S}^2(x_0,T)}.
\label{eq:calzs}
\end{align}
In the progression of linearised equations of motion for sources of arbitrary spin $L$ derived in \cite{Dietrich:2015oba,Dietrich:2016fby} it has to satisfy
\begin{align}
\Big[-T\partial_TT^{-1}\partial_T-\frac{1}{4}\frac{\Box}{T}-\frac{1}{T^2}\Big]\mathcal{S}^2=0
\end{align}
for $L=0$---which saturates the Breitenlohner-Freedman bound---for the boundary condition
\begin{align}
\lim_{T\rightarrow\varepsilon}\mathcal{S}^2(x_0,T)/T=S^2(x_0).
\end{align}
For the 4d homogeneous boundary condition $S^2(x_0)=m^2$ this has the solution
\begin{align}
\mathcal{S}^2(x_0,T)=m^2T.
\end{align}
Plugging this tachyon profile back into \eqref{eq:calzs} reproduces the factor $\E^{-m^2T}$ and thereby accommodates the effect of a constant explicit mass.

In the interacting case this profile will be modified. Also, the spin-2 background should react to this warping. Determining these effects, however, are beyond the scope of the present paper.

\section{More on the effect of warping \label{sec:warp}}

By using a similar warp factor $\E^{-cz^2}$ with a different choice of coordinates \cite{Herzog:2006ra}, 
\begin{align}
\D s^2=\frac{\D z^2/(1-z^4/z^4_h)+\D t^2(1-z^4/z^4_h)+\D\vec x^2}{z^2},
\label{eq:herzog}
\end{align}
found a sign change, which crucially depended on corrections of the type discussed in Appendix \ref{sec:matching}. Thus, the effect of this type of warping crucially depends on the choice of coordinates. [It is not the change from a coordinate of mass dimension $-1$, i.e., $z$, to one with $-2$, i.e., $T$, that makes the difference here, but that there is no singular factor on the radial AdS coordinate in \eqref{eq:patchblack}.] Worldline holography predicts the Fefferman-Graham form. 

\eqref{eq:herzog} is related to \eqref{eq:patchblack} by
\begin{align}
\frac{\D z}{z(1-z^4/z_h^4)}=\frac{\D T}{2T}.
\end{align}

~\\

Also an ad-hoc truncation of the AdS space at $T_0$ does not lead to a sign change, contrary to \cite{Herzog:2006ra}. $$\frac{\mathrm{Vol}_5-\mathrm{Vol}_5^\mathrm{BH}}{\mathrm{Vol}_3\,\beta}\propto\left\{\begin{array}{cc}\frac{1}{2T_h^2}-\frac{1}{4T_0^2}&T_0>T_h\\\frac{T_0^2}{4T_h^4}&T_0<T_h\end{array}\right.$$ In fact, in the present setting, no warp factor without sign change can lead to a phase transition, as the difference of the integrands has definite sign.

\end{document}